\documentclass[aps,prl,twocolumn,showpacs,superscriptaddress,groupedaddress]{revtex4} 

\usepackage{amsmath}
\usepackage{epsfig}
\usepackage{array}
\usepackage{verbatim}
\usepackage{hyperref}
\usepackage{amssymb}
\usepackage{multirow}

\usepackage{color}
\usepackage{ulem}

\newcommand{\beg}{\begin{equation}}

\newcommand{\en}{\end{equation}}

\newcommand {\dis}{\displaystyle}

\newcommand{\eref}[1]{Eq.~(\ref{#1})}

\renewcommand{\emph}{\textit}

\definecolor{DRed}{rgb}{0.85,0,0}

\newcommand{\beq}{\begin{equation}}

\newcommand{\va}{\varepsilon}
\newcommand{\eeq}{\end{equation}}
\newcommand{\barray}{\begin{eqnarray}}
\newcommand{\earray}{\end{eqnarray}}

\newcommand{\disp}[1]{Eq.~(\ref{#1})}
\newcommand{\refdisp}[1]{Ref.~(\onlinecite{#1})}

\newcommand{\ve}{\varepsilon}

\renewcommand{\em}{\it}

\begin{document}

\title{ Functionally independent conservations laws in a quantum integrable model}
\author{ Haile Owusu and B. Sriram Shastry}
\affiliation{ Physics Department, University of California, Santa Cruz, CA 95064, USA}

\date{ March 14, 2013}
\begin{abstract}
We study a recently proposed quantum integrable model defined on a lattice with N sites, with Fermions or  Bosons  populating each site, as a close relative of the well known spin $\frac{1}{2}$ Gaudin model. This model has $2 N$ arbitrary parameters,   a linear dependence on an interaction type parameter $x$, and can be solved exactly. It has $N$  known constants of motion that are linear in $x$.   We display further  constants of motion  with higher Fermion content that are are {\em linearly independent of the known conservation laws}.
 Our main result is that despite the existence of the higher  conservation laws,  the model has only {\em N functionally independent conservation laws}. Therefore we propose that $N$ can be viewed as the  number of degrees of freedom,  in parallel to  the classical definition of integrability. 

\end{abstract}

\maketitle

{\bf Introduction}
Quantum integrable systems have emerged from an esoteric beginning in the pioneering works of Bethe,  Onsager, Yang, Baxter \refdisp{Gaudin} and others, and are of interest to a much wider community in the recent years. Optical lattices where quantum quenches can be realized to study  dynamics away from equilibrium \refdisp{rigol} \refdisp{barmettler}, \refdisp{polkonikov}, transport   theory \refdisp{prosen} \refdisp{moore}  and considerations of entanglement entropy \refdisp{cardy} are some areas where the the standard  models of quantum integrability, such as the Ising model in a transverse field, the anisotropic Heisenberg or XXZ model, and the 1-d Hubbard model   find wide applications,  unanticipated in the pioneering studies. 

In this renewed era of interest, some basic questions about quantum integrability remained unanswered. While the Yang-Baxter equations provide a deep mathematical underpinning to this field and indeed  a source of  most of the quantum integrable models,  there are good reasons to look more broadly at the field. For instance there are other families of quantum integrable models that do not  not fit naturally into this scheme, such as the Calogero Sutherland systems.  Also, there is a well evolved language of classical integrable systems \refdisp{arnold}, with the dictionary and numerology  of ``degrees of freedom'' and  of matching  ``functionally independent conservation laws'', that is not easily translated to the quantum arena. The meaning of the  phrase {\em degrees of freedom},  is not always clear in many models of current interest.  The number of conservations laws of  even well established models, such as the XXZ model, is not quite a settled  story, since new conservation laws have been found \refdisp{prosen} in very recent studies, with important implications for the transport behavior of this model \refdisp{moore}. Discussions of the meaning of  quantum integrability is therefore of interest and has been addressed in recent literature \refdisp{shastry2},  \refdisp{Caux}, \refdisp{Zhang} and \refdisp{emil-sriram}.

In this work, we study a  recently proposed quantum lattice model \refdisp{shastry2} that is defined on a lattice of N sites, with either Fermions or Bosons, at each site, that is similar to the well known Gaudin spin $\frac{1}{2}$   model\cite{Gaudin,Dukelsky}. For the Fermion model, we show that N, the number of  lattice sites, can be viewed as the number of degrees of freedom in close analogy to the classical definition. To do this, we are able to enumerate the basic N conservation laws from the 1 particle sector.  These basic conservation laws are {\em linear in a parameter $x$}, this linearity (or simple polynomial dependence) has been highlighted in recent works \refdisp{emil-sriram}.
Our most important result is as follows:
 we find that while other non trivial (i.e. four Fermi and six Fermi)  conservation laws do exist in higher particle sector, even linearly independent from the basic ones, these are {\em functionally dependent} on the basic ones. We display the generating functions for the higher conservation laws, and the  non trivial functional relation between these.    In the case of Bosons and the Gaudin model, using an interesting algorithm, we have checked that up to at least $N=20$,  there are {\em no other non trivial conservation laws} that are polynomial in the parameter $x$, and hence the N conservation laws available from the single particle sector are again the entire set.

To summarize the models studies here let us note that  a prototypical quantum integrable model can be built up in two steps:  
{\bf (Step-I) } We  identify  a family of  real symmetric commuting matrices  $H(x)$ and $ \widetilde{H}(x)$ in N dimension, that depend linearly on a parameter $x$:
\beq
[H(x),\widetilde{H}(x)]=0, \quad \mbox{for all values of $x$}.
\label{comm}
\eeq
This problem leads to several classes of solutions that differ in the number of independent Hamiltonians $\widetilde{H}(x)$ that can be found for a given $H(x)$. These  are referred to as the Type-M with various values for M as in  \refdisp{owy} \refdisp{oy}.
Type-1 represents a maximal set of such matrices with at most $N$ commuting partners (including the identity matrix),  where all such matrices can be expressed in terms of the projection operators $\pi_{ij} = |i \rangle \langle j |$ and $3 N$ arbitrary real constants $\{ \gamma_i, \ve_i, d_i  \}$ as:
\barray
H(x)&=& \sum_i d_i Z_i \; ,   \label{eq1} \\
Z_i &=& \pi_{ii} + x \  \sum'_j \frac{ \gamma_i \gamma_j (\pi_{ij}+\pi_{ji} ) -\gamma_i^2 \pi_{jj} - \gamma_j^2 \pi_{ii}}{\ve_i-\ve_j}.  \ \ \label{eq2}
\earray
The prime   indicates the exclusion of the summed index with the fixed index.
The commutation of the $N$ basis operators  $[Z_i,Z_j]=0$ is at the heart of this construction,
and provide the $N$ ``higher constants of motion'' or dynamical symmetries, which depend parametrically on $x$. This is the maximal set of commuting operators, whose number equals the dimension of the defining space- $N$.

{\bf (Step-II)}. The next step  is to embed  the operators of \disp{eq2} into Fock space, thus $Z_i \to \hat{Z}_i$ by writing these  in terms of  quantum field operators acting upon suitable spaces. This  procedure  thereby gives rise to  a quantum integrable models with Hamiltonians $H \to \hat{H} = \sum_i d_i \ \hat{Z}_i$. We may use spins (i.e. hard core bosons), canonical  Bosons or Fermions to get models that coincide in the single particle subspace,  but in higher subspaces are  essentially different models. Let us note the Fermi model:
\beg
\hat{Z}_r \equiv n_r+x \sum_s^{'}{\dfrac{\gamma_r \gamma_s (a^\dag_r a_s + a^\dag_s a_r )-\gamma_r^2 n_s- \gamma_s^2 n_r}{\varepsilon_r-\varepsilon_s}},
\label{Z's}
\en
where $\{a_i,a^\dag_j\}=\delta_{ij}$, $n_r = a^\dag_r a_r$, and   $[\hat{Z}_r,\hat{Z}_s]=0$ for all $r,s=1,2,\dots,N$ and for all values of $x$. Different commuting operators, linearly dependent on the parameter $x$, are expressible as in \disp{eq1}. With spins in place of Fermions, this  model reduces to the well known Gaudin model, generalized to include the coefficients $\gamma_i$, and with (canonical)  Bosons  we get yet another model. 
 
  In every higher particle number  sectors, these operators for any choice of the embedding Fock space have obvious  descendent representatives that commute with each other and with $\hat{H}$. 
For instance in the 2 particle sector for Fermions or the  spin-$\frac{1}{2}$ Gaudin model, the Hilbert space has dimension of $\binom{N}{2}$, where the above operators provide $N$ commuting operators.     These  are of limited interest to us. Our aim is to seek out {\em  operators that are linearly  independent} of these descendent operators.  In this larger  space we could for instance,  independently construct type-1 matrices that are $\binom{N}{2}$ in number, 
 -- ostensibly there's room for additional, as yet undiscovered conservation quantities that act in the two-particle sector but are null in the one-. 

In fact our main goal in this work is to examine the proposal that $N$ is the maximum number of {\em functionally independent commuting operators} for $\hat{H}$. Our  search for functional independence starts with the more restrictive but technically feasible search for  linear independence, followed by the test of functional dependence.

{\bf Algorithm for Finding Additional Conservation Laws:} To uncover any such additional conservation laws, one can use an algorithm that we next describe. This algorithm can be implemented numerically for a relatively small size of the Hilbert space  ${\cal N}$, as an illustration  for Fermions  ${\cal N} \sim \ \binom{N}{r}$ where $r$ is the number of particles.   We used ${\cal N} \leq  \binom{10}{3}$   and  in our studies of the Fermi, Bose and spin-$\frac{1}{2}$ Gaudin models for , and the results are described below.

The algorithm  uses the matrix decomposition outlined in \refdisp{shastry2} and \refdisp{owy} and the current-constructing algorithm outlined in \cite{oy}. In ${\cal N}$ dimensions,  if $[H(x),\widetilde{H}(x)]=0$, it can be shown that there exists an antisymmetric matrix $S$ such that
\begin{equation}
\begin{array}{l}
\dis H(x)=T+x\Big([T,S]+W\Big)\\
\\
\dis \widetilde{H}(x)=\widetilde{T}+x\left([\widetilde{T},S]+\widetilde{W}\right)\\
\end{array}
\label{S's}
\end{equation}
where $T,\widetilde{T},W,\widetilde{W}$ all mutually commute and, without loss of generality, we take them to be diagonal matrices. Given this decomposition \eref{comm} is satisfied if
\beg
\left[[T,S], [\widetilde{T},S] \right]+\left[W,[\widetilde{T},S]  \right]-\left[\widetilde{W},[T,S]  \right]=0
\en
For a given $H(x)$, $S,T$ and $W$ can be determined. Finding a conserved current $\widetilde{H}(u)$, i.e. finding a solution for the non-zero elements of diagonal matrices $\widetilde{T}$ and $\widetilde{W}$, is then a matter of solving $\binom{{\cal N}}{2}$ simultaneous linear equations in $2{\cal N}$ unknowns, where the number of linearly independent solutions is the number of independent conservation laws \cite{expl}.

{\bf Fermions: The Four Fermi Conserved Current}

We applied the above algorithm to the matrices arising from the action of the Type 1 Fermionic  Hamiltonian  $\hat{H}= \sum_j d_j \hat{Z}_j$ \disp{Z's}, in the two particle sector and found that there is a $2N-3$ member family of mutually commuting currents. $N$ of these are just the action of the original currents $\hat{Z}_r$ (see \eref{Z's}) in that sector. The $N-3$ other conserved quantities correspond to the particle action of \textit{four} Fermi operators with a generating function of the  form:
\begin{equation}
\begin{split}
\hat{Q}(\alpha)= \frac{1}{2} \sum'_{i j} &\frac{n_{i}n_j}{(\alpha- \varepsilon_i)(\alpha-\varepsilon_j)} \\
+& \frac{1}{2} x \sum'_{i,j,k} \frac{\phi_{ijk}}{(\alpha- \varepsilon_i)(\alpha-\varepsilon_j)(\alpha-\varepsilon_k)},
\end{split}
\label{q-def}
\end{equation}
where $\phi_{ijk} \equiv \omega_{ij} n_k + n_i n_j \gamma_k^2$ and $\omega_{ij}\equiv \gamma_i \gamma_j (a^\dag_i a_j+a^\dag_j a_i) - n_i \gamma_j^2 - n_j \gamma_i^2$. 
Operators analogous to \disp{Z's} that are indexed by ``r'' can be found by taking residues
\beq \hat{Q}_r = \lim_{\alpha \to \va_r} \ (\alpha - \va_r) \ Q(\alpha), \label{qrs}
\eeq
these are seen to be  (i) linear in $x$ (ii) are $N-3$ in number and (iii) linearly independent of $\hat{Z}_r$.  The result (ii) requires an elaborate proof of three relationships between the apparently independent $\hat{Q}_r$.   In particular one can show that $\sum_i{\hat{Q}_i}=0$, and we skip the remaining two  for brevity.

{\bf Analytical Proof of commutation:}
In what follows we will prove that the  generating function  $\hat{Q}(\alpha)$ is  conserved  i.e. 
\beg
[\hat{Z}_r, \hat{Q}(\alpha)]=0
\label{qcomm}
\en
for all $r=1,2,\dots,N$ and all values of $x$ and $\alpha$. Towards this end it is useful to define the quantity $\beta_{ij} \equiv a^\dag_i a_j-a^\dag_j a_i$

The commutator has  terms of $O(x^n)$ with $n=0,1,2$ and each must vanish identically.
The $O(x^0)$ terms vanish as all number operators mutually commute.

To satisfy \eref{qcomm} to $O(x)$ we must show \cite{expl-comm} that
\beg
\begin{split} 
\sum'_{j k} 
[   \frac{w_{ij}}{\varepsilon_i - \varepsilon_j},  \frac{n_i n_k}{(\alpha- \varepsilon_i)(\alpha-\varepsilon_k)}+\frac{n_j n_k}{(\alpha- \varepsilon_j)(\alpha-\varepsilon_k)}]\\
= \sum'_{j,k} [ \frac{\phi_{ijk}}{(\alpha- \varepsilon_i)(\alpha-\varepsilon_j)(\alpha-\varepsilon_k)}, n_i], 
\label{first}
\end{split}
\en
where we will subsequently follow the convention in which the prime on summation implies that all indicated indices are distinct. The  RHS evaluates to
 $$  - \sum'_{j,k}  \frac{   \gamma_i \gamma_j \beta_{ij} n_k }{(\alpha- \varepsilon_i)(\alpha-\varepsilon_j)(\alpha-\varepsilon_k)}$$
 In the LHS, we note that $[w_{ij},n_i n_k] = - [w_{ij} , n_j n_k]= - \gamma_i \gamma_j \beta_{ij} n_k$  and using the partial fraction identity
 
 \beg
 \frac{1}{\varepsilon_i -\varepsilon_j} \ ( \frac{1}{\alpha -\varepsilon_i} - \frac{1}{\alpha -\varepsilon_j})= \frac{1}{\alpha -\varepsilon_i} \times  \frac{1}{\alpha -\varepsilon_j}, \label{partial-frac}
 \en
 we see that \eref{first} is satisfied.
 
 To $O(x^2)$ we need to show
 \beg
 \sum'_j \frac{[ w_{ij},  \hat{Q}'(\alpha)]}{\varepsilon_i - \varepsilon_j}=0, \label{ox2}
 \en
 where $\hat{Q}'(\alpha) =\frac{d}{d x} \hat{Q}(\alpha)$. We see from \eref{q-def} that $\hat{Q}'$ depends upon three indices and so for the  fixed pair in  $w_{i j}$ in the \eref{ox2}, we may organize $\hat{Q}'$ as follows. The terms not involving either $i$ or $j$ commute and can be neglected. We may have both these  indices in a class of terms  $\hat{Q}'$, or only one index in another class of terms. Let us look at them separately:
 
 {\bf Two indices common:}
 The relevant part of $\hat{Q}'$ in \eref{ox2} may then be written   as
 $$\sum'_k \frac{1}{(\alpha- \varepsilon_i)(\alpha-\varepsilon_j)(\alpha-\varepsilon_k)} \ \left( \phi_{ijk} + \phi_{kij}+ \phi_{jki}\right) $$
 Now $w_{ij}$ commutes with the first of these terms and for the others we calculate
 \beg
 [w_{ij},\phi_{kij}]= -(\gamma_i^2+\gamma_j^2) \ \beta_{ij} n_k + \gamma_i \gamma_k \beta_{jk} n_i  - \gamma_j \gamma_k \beta_{ik} n_j, 
 \label{e14}
 \en
 and the term $[w_{ij}, \phi_{jki}]$ is obtained by exchanging $i \leftrightarrow j$ in \eref{e14}, giving the negative of the first so that the sum  vanishes identically. 
 
 {\bf One index common:}
 The relevant terms in \eref{ox2}  may be written as
 \beg
 \sum'_{j, k<l} [\frac{w_{ij}}{\varepsilon_i-\varepsilon_j}, \left( \frac{\phi_{ikl}+\phi_{ilk}+\phi_{kli} }{(\alpha- \varepsilon_i)(\alpha-\varepsilon_k)(\alpha-\varepsilon_l)} + ( i \to j) \right)], \label{eq20}
 \en
  where we have used the symmetry of $\phi_{ijk}$ in the first two indices to isolate all terms with a triple of indices $ilk$ that have one index  $i$.
  It is easy to see that
  \begin{align}
&[w_{ij}, \phi_{kli}]= - \gamma_i \gamma_j \ \beta_{ij} w_{kl} \label{eq18} \\
~&[w_{ij}, \phi_{ikl}+\phi_{ilk}]= - \gamma_i \gamma_j \gamma_k \left(\beta_{ik} \gamma_j - \beta_{jk} \gamma_i \right) n_l- (k \leftrightarrow l) \label{eq19} \\
~&[w_{ij}, \phi_{kli} \phi_{ikl}+\phi_{ilk}]= \gamma_i \gamma_j M_{ijkl}.
  \end{align} 
  Here $\gamma_i \gamma_j M_{ijkl}$ is defined as the sum of the two terms \eref{eq18} and \eref{eq19}. It   is clearly symmetric in $k \leftrightarrow l$  and also  antisymmetric in  exchanging $i \leftrightarrow j$, so that using \eref{partial-frac}  we can rewrite  the LHS of \eref{ox2} as
 \beg
 \sum'_{j,  k<l} \frac{\gamma_i \gamma_j M_{i j k l} }{(\alpha- \varepsilon_i)(\alpha- \varepsilon_j)(\alpha-\varepsilon_k)(\alpha-\varepsilon_l)}
 \en 
  We  note an identity satisfied by cyclically permuting the indices $j k l$:
  \beg
  \gamma_j M_{i jkl}+ \gamma_k M_{i klj}+ \gamma_l M_{i ljk} =0, 
  \en
  whereby the commutator vanishes identically.

{\bf  Functional relationship between the Generating Functionals:}
Let us note the generating function for the two Fermi currents:
\beg
\hat{Z}(\alpha) = \sum_i \frac{n_i}{\alpha- \varepsilon_i} + \frac{1}{2} x \sum'_{i j} \frac{w_{ij}}{(\alpha- \varepsilon_i)(\alpha-\varepsilon_j)}, \label{zgen}
\en
so that $[\hat{Z}(\alpha), \hat{Z}(\beta) ]=0$ for any $\alpha,\beta$, and 
\beq
\hat{Z}_i = \lim_{\alpha \to \varepsilon_i} (\alpha - \varepsilon_i) Z(\alpha). \label{zrs}
\eeq
This is similar to the generating function for the four Fermi currents in \disp{q-def}. For completeness we note that from \disp{qrs}
\begin{align}
\hat{Q}_i &=&  \sum'_j \frac{ n_i n_j}{\varepsilon_i-\varepsilon_j} + x \sum'_{j,k} \frac{\phi_{ijk} + \phi_{kji}}{(\varepsilon_i-\varepsilon_j)(\varepsilon_i-\varepsilon_k)}.
\end{align}
To summarize, we computed a bilinear  generating function in \disp{zgen} and showed that it has a four Fermionic set of constants of motion $\hat{Q}_r$'s in \disp{qrs}, with a generating function \disp{q-def}, that are linear in $x$ and also   linearly independent of the $\hat{Z}_r$ in \disp{zrs}. 
  Remarkably these four-Fermi currents, while linearly independent of the $\hat{Z}$'s are not \textit{functionally} independent of them. Indeed a brief calculation shows   that
\beg
\hat{Q}(\alpha) = \frac{1}{2} \left[ \dfrac{\hat{Z}(\alpha)^2}{1-\frac{x}{\tilde{x}(\alpha)}}+\left( 1-\frac{x}{\tilde{x}(\alpha)} \right) \frac{d}{d \alpha} \left( \dfrac{\hat{Z}(\alpha)}{1-\frac{x}{\tilde{x}(\alpha)}}\right) \right]
\label{funct}
\en
where $\tilde{x}(\alpha) \equiv \left( \sum_m{\gamma_m^2/(\alpha-\varepsilon_m)} \right)^{-1}$. 

To acquire some feeling for these generating functions, it is useful to transform to the diagonal representation for the  quadratic Fermion model. In \cite{shastry2}  one finds modes that  diagonalize all the $\hat{Z}_i$ and, by extension $\hat{Z}(\alpha)$ by constructing the x-dependent canonical fermion creation/annhilation operator set:
\beg
d^\dag_i=\sum_j{\dfrac{\phi_i \gamma_j}{\lambda_i(x)-\varepsilon_j}a^\dag_j}, \quad \mbox{with} \quad \{d_i,d^\dag_j \}=\delta_{ij},
\en
where we require that $\phi_i^{-2}= \sum_j{\left( \gamma_j/(\lambda_i(x)-\varepsilon_j) \right)^2}$ and $\sum_m{\gamma_m^2/(\lambda_i(x)-\varepsilon_m)}=1/x$. It follows that  the two-Fermi generating function can expressed as 
\beg
\hat{Z}(\alpha) =\left(1-\frac{x}{\tilde{x}(\alpha)} \right)\sum_m{\dfrac{\tilde{n}_m}{\alpha-\lambda_m(x)}},
\en
where $\tilde{n}_m \equiv d^\dag_m d_m$, $x$-dependent fermionic number operator. \eref{funct}  can be rewritten in terms of these operators as:
\beg
\hat{Q}(\alpha) =\left(1-\frac{x}{\tilde{x}(\alpha)} \right)\sum_{i<j}{\dfrac{\tilde{n}_i \tilde{n}_j}{\left( \alpha-\lambda_i(x) \right) \left( \alpha-\lambda_j(x) \right)}}.
\en
In this representation, since both generating functions are constituted of the number operators in the diagonal basis, their mutual  commutation is evident. However, known the linear dependence of the two generating functions on the parameter $x$ is now  hidden.

{\bf Higher order Fermionic conservation laws:}
We can continue the logic of the above section to higher particle sectors. While a more complete discussion is possible, we shall content ourselves with a short presentation of our results. The Type 1 Fermionic model  appears to possess a  hierarchy of conservation laws, e.g. four-, six-,eight-Fermi terms, etc which are linear in $x$. We find from the above  algorithm
that a sequence of generating functions exist with $m=1,2,3,\ldots$ where
\beg
\hat{C}^{(m)}(\alpha) \equiv \frac{1}{m!} \left(1-\frac{x}{\tilde{x}(\alpha)} \right)\sum'_{k_1,k_2,\dots,k_m}{\prod_j^m{\dfrac{\tilde{n}_{k_j}}{\alpha-\lambda_{k_j}(x)}}},
\label{gens}
\en
is a $2m$-Fermi conserved current also linear in $x$, for $m < N$. Thus $\hat{Z}$ and $\hat{Q}$ discussed above correspond to $m=1,2$. {If we denote reduced generating functions by $\hat{C}^{(m)}(\alpha)= (1- \frac{x}{\hat{x}(\alpha)}) \  \hat{C}_{red}^{(m)}(\alpha) $, then these satisfy the differential equation
\beq
 \hat{C}_{red}^{(m)}(\alpha) = \frac{1}{m} \left( \frac{d}{d \alpha} +  \hat{C}_{red}^{(1)}(\alpha) \right)  \hat{C}_{red}^{(m-1)}(\alpha). 
\label{reds}
\eeq

Though the commutativity of the generating functions is a straightforward consequence of \eref{gens}, deriving the linearity in $x$ of $\hat{C}^{(m)}(\alpha)$  is a bit more involved. In the (unrotated) original representation  { the linear $x$-dependence can be made explicit, i.e.} 
\begin{equation}
\begin{split}
\hat{C}&^{(m)}(\alpha) =\\
&\sum'_{k_1,k_2,\dots,k_m} \left[\underbrace{ \dfrac{1}{m!}\prod_j^m{\dfrac{n_{k_j}}{\alpha-\varepsilon_{k_j}}}\left(1-x\sum_{s>m}{\dfrac{\gamma_{k_s}^2}{\alpha-\varepsilon_{k_s}}} \right) }_{c^{(m)}_1} \right.\\
&+\left. \underbrace{ x\, \dfrac{1}{2(m-1)!} \sum_{s>m}  \gamma_{k_s} \gamma_{k_m}  \dfrac{ a_{k_s}^\dag a_{k_m}+a_{k_m}^\dag a_{k_s} }{(\alpha-\varepsilon_{k_s})(\alpha-\varepsilon_{k_m})}\prod_j^{m-1}{\dfrac{n_{k_j}}{\alpha-\varepsilon_{k_j}}} }_{c^{(m)}_2} \right],
\end{split} \label{linx}
\end{equation}
where $k_1,k_2,\dots,k_N$ is an arbitrary permutation of the list of integers from $1$ to $N$.
For example, the six-fermi conserved current is:
\begin{equation}
\begin{split}
\hat{C}&^{(3)}(\alpha) =\\
&\sum'_{k_1,k_2,k_3} \left[\dfrac{1}{6}\dfrac{n_{k_1}n_{k_2}n_{k_3}}{(\alpha-\varepsilon_{k_1})(\alpha-\varepsilon_{k_2})(\alpha-\varepsilon_{k_3})}\left(1-x\sum_{s>3}{\dfrac{\gamma_{k_s}^2}{\alpha-\varepsilon_{k_s}}} \right) \right.\\
&+\left. x\, \dfrac{1}{4} \sum_{s>3} \dfrac{n_{k_1}n_{k_2} \left(a_{k_3}^\dag a_{k_s}+ a_{k_s}^\dag a_{k_3} \right)}{(\alpha-\varepsilon_{k_1})(\alpha-\varepsilon_{k_2})(\alpha-\varepsilon_{k_3})(\alpha-\varepsilon_{k_s})} \right].
\end{split}
\end{equation}

The proof proceeds by induction. Assume \eref{linx} true for $\hat{C}^{(m-1)}(\alpha)$. Note that the product $(1-x/\tilde{x}(\alpha))^{-1} \hat{C}^{(1)}(\alpha) \cdot \hat{C}^{(m-1)}(\alpha)$ contains $(m-1)$-Fermi terms, as well as $m$-Fermi -- we are concerned with the latter, as the former are exactly cancelled in \eref{reds} by the $\frac{d}{d \alpha} \hat{C}^{(m-1)}(\alpha)$ term. The $m$-Fermi terms are given by

\begin{equation}
\begin{split}
\left[ c^{(1)}_1 c^{(m-1)}_1\right]_{m} &= \\
\sum'_{k_1,k_2,\dots,k_m}&{\left[ m \left( 1-\frac{x}{\tilde{x}(\alpha)} \right)^2 +x \,\sum_i^m{\dfrac{\gamma_{k_i}^2}{\alpha-\varepsilon_{k_i}}} \left( 1-\frac{x}{\tilde{x}(\alpha)} \right) \right.}\\
+ 2x^2 \, \sum'_{i,j}&\left.{\dfrac{\gamma_{k_i}^2 \gamma_{k_j}^2}{(\alpha-\varepsilon_{k_i})(\alpha-\varepsilon_{k_j})}} \right] \dfrac{1}{m!}\prod_j^m{\dfrac{n_{k_j}}{\alpha-\varepsilon_{k_j}}}\\
\end{split}
\end{equation}

\begin{equation}
\begin{split}
&\left[ c^{(1)}_1 c^{(m-1)}_2 +c^{(1)}_2 c^{(m-1)}_1 \right]_{m} = \\
&\qquad \qquad \qquad \qquad \left[ m  \left( 1-\frac{x}{\tilde{x}(\alpha)} \right) +2x\sum_i^m{\dfrac{\gamma_{k_i}^2}{\alpha-\varepsilon_{k_i}}}\right]\cdot\\
&x\, \dfrac{1}{2(m-1)!} \sum_{s>m}  \gamma_{k_s} \gamma_{k_m}  \dfrac{ a_{k_s}^\dag a_{k_m}+a_{k_m}^\dag a_{k_s} }{(\alpha-\varepsilon_{k_s})(\alpha-\varepsilon_{k_m})}\prod_j^{m-1}{\dfrac{n_{k_j}}{\alpha-\varepsilon_{k_j}}}\\
\end{split}
\end{equation}

\begin{equation}
\begin{split}
&\left[ c^{(1)}_2 c^{(m-1)}_2\right]_{m} = \\
&-2x^2\sum'_{k_1,k_2,\dots,k_m} \left[ \sum'_{i,j}{\dfrac{\gamma_{k_i}^2 \gamma_{k_j}^2}{(\alpha-\varepsilon_{k_i})(\alpha-\varepsilon_{k_j})}} \dfrac{1}{m!}\prod_j^m{\dfrac{n_{k_j}}{\alpha-\varepsilon_{k_j}}} \right.\\
&+d_m \sum_i^m{\dfrac{\gamma_{k_i}^2}{\alpha-\varepsilon_{k_i}}}  \sum_{s>m}  \gamma_{k_s} \gamma_{k_m}  \dfrac{ a_{k_s}^\dag a_{k_m}+a_{k_m}^\dag a_{k_s} }{(\alpha-\varepsilon_{k_s})(\alpha-\varepsilon_{k_m})}\prod_j^{m-1}{\dfrac{n_{k_j}}{\alpha-\varepsilon_{k_j}}}\\
\end{split}
\end{equation}
where $d_m=(2(m-1)!)^{-1}$. Summing these terms yields a $(1-x/\tilde{x}(\alpha))^{-1} \hat{C}^{(1)}(\alpha) \cdot \hat{C}^{(m-1)}(\alpha)$ linear in $x$ and, thus, $\hat{C}^{(m)}(\alpha)$ is also linear in $x$.

{\bf Bosonic and spin models}.
We have used the above algorithm to study the higher particle number sectors for Bosons  on a lattice with $N \leq 6$ and the number of particles $2 \leq  n \leq 6$  and find that {\em there are no linear conservation laws beyond the $\hat{Z}_j$  that are linear in $x$.} For the spin-$\frac{1}{2}$ Gaudin model, we have studied higher particle number sectors with  arbitrary $\gamma_i$, where the  operators  analogous to $Z_j$   do exist. In that case we find that again {\em there are no other conservation laws that are linear in $x$}.

{\bf Discussion}
The Type 1 fermionic model appears to be unique in possessing these `extra' conservation laws linear in the parameter $x$. Not only do the Type 1 bosonic and Gaudin models not have them, numerical experiments with randomly generated Type $M$ -- $M>1$ -- fermionic and bosonic models similarly exhibit no such additional conservation laws. As we have shown, however, these extra Type 1 conservation laws are functionally dependent on the $N$ `original' ones -- the $\hat{Z}_j$. 

In conclusion, we have provided a non trivial  example of the enumeration of dynamical symmetries  in the model considered. It appears the be the first  case where one can establish the {\em functional dependence} between a ``higher set'' of conservation laws on a more fundamental set, surprisingly both being linear in the parameter $x$. With this, we have a close parallel to the classical idea of integrability, and a well defined enumeration of {\em the degrees of freedom} that have so far been missing.
  
{\bf Acknowledgements:}
 This work was supported by DOE under Grant No. FG02-06ER46319.


\begin{thebibliography}{99}
\bibitem{expl} It should be emphasized that this algorithm yields trivial $\widetilde{H}(x)$ for general $H(x)$. The algorithm will only yield nontrivial results for an integrable matrix $H(x)$.
\bibitem{expl-comm} We have used the identity $[w_{ij},n_i n_j]=0$ and also noted that on the right hand side, the density $n_i$ commutes with the variables labelled by $k$ in $\phi_{i,j,k}$
\end{thebibliography}
\end{document}